\documentclass[twocolumn, amsmath,amssymb]{revtex4}

\usepackage{graphicx}
\usepackage{subfigure}
\usepackage{dcolumn}
\usepackage{bm}
\usepackage{threeparttable}
\usepackage{pslatex}
\usepackage{color}

\newcommand{\be}{\begin{equation}}
\newcommand{\ee}{\end{equation}}
\newcommand{\syn}{synchronization}
\newcommand{\xo}{x_i[t]}
\newcommand{\xt}{x_j[t]}
\newcommand{\cc}{cross correlation}
\newcommand{\pv}{$p$-value}
\newcommand{\ex}{excitatory}

\begin{document}

\title{Network inference --- with confidence --- from multivariate time series}

\author{Mark A. Kramer}
\altaffiliation{Department of Mathematics and Statistics, Boston University, Boston, MA, 02215, USA.  Email:  mak@bu.edu}
\author{Uri T. Eden$^*$}
\author{Sydney S. Cash}
\altaffiliation{Epilepsy Service, Department of Neurology, Harvard Medical School, ACC 835, Massachusetts General Hospital, 55 Fruit Street, Boston, MA, 02114, USA}  
\author{Eric D. Kolaczyk$^*$}

\begin{abstract}
Networks --- collections of interacting elements or nodes --- abound in the 
natural and manmade worlds.  For many networks, complex spatiotemporal dynamics  stem from patterns of physical interactions unknown to us.  To infer these interactions, it is common to include edges between those nodes whose time series exhibit sufficient functional connectivity, typically defined as a measure of coupling exceeding a pre-determined threshold.  However, when uncertainty exists in the original network measurements, uncertainty in the inferred network is likely, and hence a statistical propagation-of-error is needed.  In this manuscript, we describe a principled and systematic procedure for the inference of functional connectivity networks from multivariate time series data.  Our procedure yields as output both the inferred network and a quantification of uncertainty of the most fundamental interest:  uncertainty in the number of edges.  To illustrate this approach, we apply our procedure to simulated data and electrocorticogram data recorded from a human subject during an epileptic seizure. We demonstrate that the procedure is accurate and robust in both the determination of edges and the reporting of uncertainty 
associated with that determination.
\end{abstract}

\maketitle

\section{Introduction
\label{sec:intro}}
\begin{figure}[tbh]
\centering
\includegraphics[angle=0, height=2.5cm, width=8.6cm]{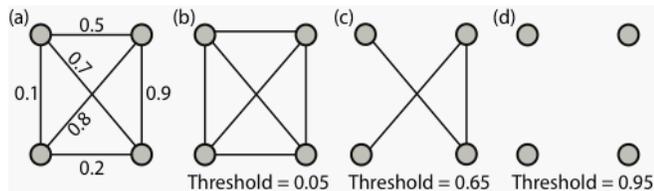}
  \caption{\label{fig:ex}  The number of edges in this 4-node network depends upon our choice of coupling threshold.  Each node (gray circle) represents a spatial location from which we record time series data.  Applying a coupling measure to the dynamic activity recorded at node pairs, we obtain the normalized (between $0$ and $1$) coupling values shown in (a).  If we choose the coupling threshold too low (0.05) we include edges between all nodes as in (b).  If we choose the coupling threshold too high (0.95) we obtain no edges in the network as in (d).  An intermediate choice of coupling threshold (0.65 in c) yields a different network.}
\end{figure}

Many examples of natural and fabricated networks exist in the world, including airline networks, computer networks, and neural networks.  To define a network is, in principle, straightforward: we simply identify a collection of nodes and edges \cite{Wasserman:1994qy, Newman:2003p10236, Boccaletti:2006lr}.  A node (or vertex) represents a participant or actor in a network, while an edge represents a link or association between two nodes (Fig.\ \ref{fig:ex}).  For example, in an airline network, individual airports constitute nodes and direct connections between airports identify edges.  In a neural network, individual neurons and the physical connections between neurons determine the nodes and edges of the network, respectively.  Having established network representations of these complex systems, we may then address pertinent issues, such as the worldwide spread of infectious disease through the airline network \cite{Hufnagel:2004p10828} or the effect of cortical lesions on brain dynamics \cite{Honey:2008p8323}.

The decision to link two nodes with an edge varies in difficulty.  In some cases, a known physical connection exists between two nodes, and the choice to include an edge is then obvious.  Does an airline connect two cities or not \cite{Amaral:2000rt}?  Do two actors collaborate on a film or not \cite{Watts:1998vn,Newman:2001p10407}?  Does a physical connection exist between two brain regions or not \cite{White:1986p10157, Marder:2007p219}?  In these cases, the decision to include a link between nodes is simple and based on the known association or physical connection between two nodes.

In other cases, the interactions between nodes are obscure.  For example, we may only observe the dynamic activity at individual nodes and have no access to the physical connections between nodes.  In these cases, we may apply coupling measures to multivariate time series data associated with the node dynamics and attempt to infer their associations, without explicit knowledge of their structural connections \cite{Yu:2006p10263, Timme:2007p10300, Napoletani:2008p10264}.  This approach has proven useful in, for example, climate studies \cite{Gozolchiani:2008p10406, Yamasaki:2008p10391} and also human brain studies, in which the structural connections between brain regions remain difficult to classify (although perhaps not for long \cite{Hagmann:2008p8326}).

Viewed from a statistical perspective, two key challenges are inherent in this task of network inference: (i) appropriate interpretation of the coupling results in declaring network edges, and (ii) accurate quantification of the uncertainty associated with the resulting network.  The simplest -- and, indeed, most common -- method to interpret the coupling results and declare network edges involves comparison of the coupling strength to a threshold value \cite{Stam:2004p9785, Micheloyannis:2006p9787,Ponten:2007p4558,Srinivas:2007p4385,Stam:2007p8804, Kramer:2008p10158, Supekar:2008lr,Yamasaki:2008p10391}.  If the coupling strength between two nodes exceeds this threshold, then we connect these nodes with an edge;  otherwise, we leave the nodes unconnected.  The number of edges in the resulting network depends critically on the choice of coupling threshold (as we illustrate schematically in Fig.\ \ref{fig:ex}).  Furthermore, for a given choice of threshold, we expect a certain rate of error in (mis)declaring the connectivity status between pairs of nodes.  This network uncertainty ---intimately tied to choice of threshold --- is often overlooked.

How do we choose such a threshold?  One strategy is to apply a variety of different thresholds and examine the resulting collection of networks for robustness as a function of threshold \cite{Micheloyannis:2006p9787, Stam:2007p8804, Supekar:2008lr}.  This procedure of redundant analysis --- which some of these authors have recently employed \cite{Kramer:2008p10158} --- is both time consuming and unsatisfying.  Instead, a threshold should be chosen in a principled way, for example one that links the choice of threshold with the achievement of a pre-specified level of network uncertainty.  Such is the goal of this paper.  In referring to ``network uncertainty" many aspects of the network structure might be of interest (e.g., connectivity, degree distribution, or clustering).  Here we will focus on the most basic aspect of network confidence:  the presence or absence of network edges.  Our particular goal is to equip the process of threshold-based inference of a network with a number quantifying the expected rate of falsely declared network edges.  This number serves as a natural measure of network confidence.

In what follows, we adopt a statistical hypothesis testing paradigm to analyze multivariate time series data and create network representations of functional connectivity.  The general paradigm involves three steps:  1) calculate the strength of coupling between time series data recorded at node pairs, 2) threshold each coupling measure through the use of a formal statistical hypothesis test, and 3) control the rate of falsely declared edges through the use of statistical multiple testing procedures.  In Section \ref{sec:general}, we present a high-level outline of this general protocol, while in Section \ref{sec:ex}, we develop the procedure in detail, making specific choices of methodology for each step.  We apply the protocol to three data sets in Section \ref{sec:res} and show, in particular, that appropriate handling of the significance tests is vital.  Fundamentally, the proposed protocol is a way of constructing functional networks that, rather than emerging as the result of some arbitrarily chosen coupling threshold, are composed of edges selected to achieve a guaranteed level of overall network accuracy.  That is, it is a way of constructing networks with confidence.

\section{General paradigm
\label{sec:general}}

In this paper we are interested in the inference of networks (or, more precisely, graphs) 
$G=(V,E)$ in which edges $\{i,j\}\in E$ indicate a coupling (perhaps at nonzero lag) between time series $x_i[t]$ and $x_j[t]$ observed at $N$ nodes $i,j \in V$.  Our primary motivation is the desire to infer networks reflecting the functional (as opposed to structural) topology of neural systems.  This goal is reflected in our terminology, as well as in the numerical illustrations we present in Section \ref{sec:res}.  However, the methods we propose --- and the underlying principles upon which we base the methods --- have quite general applicability.

Network inference problems come in many varieties.  See Chapter 7 of \cite{Kolaczyk:2009fk} for a recent overview of this highly active area.  The type of networks we wish to infer are commonly called association networks.  Broadly speaking, most methods proposed to date for the inference of such networks assume independent measurements at each node.  A primary example of this paradigm is the popular problem of inferring Gaussian graphical models.  Methods for doing so include
classical methods of maximum-likelihood-based testing (e.g., see Chapter 6 of \cite{Whittaker:1990fk} or Chapter 5 of \cite{Lauritzen:1996uq}), and more recent methods based on multiple testing (e.g., \cite{Drton:2004p10935, Schafer:2005p10936,Schafer:2005p10939}) and sparse statistical inference (e.g., \cite{Meinshausen:2006kx, Friedman:2008p10941}).

However, work of this sort invariably assumes independent measurements in time rather than temporally correlated time series of interest here.  And furthermore, most of these methods (e.g., classical and those based on sparse inference principles)  are not aimed at providing a quantification of uncertainty with the inferred network.  Alternatively, there is also a sub-literature on the inference of association networks from temporally correlated data (e.g., \cite{Micheloyannis:2006p9787, Stam:2007p8804, Supekar:2008lr, Kramer:2008p10158}).  But the quantification of network uncertainty does not seem to have received much attention there.

We implement a procedure to create functional topologies from multivariate time series data that involves three general steps.  First, a coupling measure --- here, the cross correlation --- is specified and applied to the data, yielding a noisy indication of the functional connectivity between all nodes pairs.  In the neurological data described below, this measure captures the extent of interactions between 
activity recorded simultaneously at separate spatial locations of the brain.  
Second, we develop significance tests appropriate for our choice of coupling measure, and associate a statistical \pv\ with each coupling result.  Third, we analyze the resulting \pv s using principles 
of statistical multiple testing to construct a network representation of the 
functional connectivity.  In the course of this last step, we determine a 
number controlling the proportion of falsely inferred edges.  We present
this number as a basic and natural measure of network uncertainty.

Of course, the first step above implements a version of the standard approach to constructing such functional networks.  In neuroscience, for example, investigators (including some of these authors) typically specify a measure of coupling and then assign edges between node pairs whose coupling is judged to be sufficiently strong.  However, determination of just how strong is strong enough is invariably {\it ad hoc} or, at best, driven by ``expert judgement".  As a result, there is no way to annotate the resulting networks with any indication of their inherent (in)accuracy. The subsequent steps in the proposed approach, therefore, are critically important to produce networks accompanied by accurate characterizations of their uncertainty.  Put another way, we are interested here in the propagation of uncertainty in network inference, from the original time series data $x_i[t]$ to a final assessment of network 
uncertainty.  Our statistical hypothesis testing procedure, described above in three steps, achieves this goal.  Furthermore, our numerical results indicate that it in fact does so in a robust fashion.

We achieve our goal primarily through careful attention to the interdependency among each of the three steps.  In so doing, we also demonstrate how lack of such attention
can lead to nonsensical network uncertainty statements.
For example, the choice of coupling measure in the first step affects the hypotheses 
tested in the second step (i.e., the null hypothesis $H_0:$ {\em No Coupling}, 
versus $H_1:$ {\em Coupling}).  The declaration of either edge or non-edge for each pair of nodes $i,j\in V$ corresponds to either rejection of the null 
hypothesis or a failure to do so, respectively.  If rejection is determined by 
comparison of the observed coupling values to a threshold, clearly the choice 
of threshold will affect the network results.  But if we also wish to 
propagate uncertainty --- from the original time series data, through 
the testing procedure, to the final
network inferred -- it is necessary to construct accurate probabilistic
statements appropriate for the particular coupling measure we choose.

In constructing functional networks, we must consider the collection of individual hypothesis tests as a whole.  We note that the classical approach to calibrating individual hypothesis tests is not appropriate here.
If our hypothesis tests are conducted at some significance level $\alpha$
then, for each pair of nodes $i,j\in V$, we expect an edge to be declared
falsely between them with probability $\alpha$.  However, since there
are $N(N-1)/2$ such tests to be conducted (assuming an undirected network
of $N$ nodes), we actually expect $\alpha [N(N-1)/2]$ edges to be declared 
falsely over the network as a whole, assuming independence of tests.  This suggests that as the network size increases we must decrease $\alpha$ to limit the
total number of falsely declared edges.  But this strategy in turn has
the undesirable effect of decreasing the statistical power with which
we can detect edges.  This is the so-called ``multiple testing problem" in 
statistics.

Alternatively, therefore, we instead focus upon controlling the {\em rate}
of falsely declared edges.  Conditional on at least one edge being 
declared, the expected proportion of falsely declared edges here is equivalent 
to what is called the {\it false discovery rate (FDR)} in the statistical 
literature.   The control of FDR in multiple testing situations, ranging from 
signal and image processing to genome-wide association testing, has become a 
{\it de facto} standard technique for addressing the multiple testing problem (see~\cite{Dudoit:2008uq}, for example).  In fact, the use of FDR control occurs with increasing frequency in the network literature as well (e.g., \cite{Faith:2007p10833,Nariai:2007p10832}).  However, there is little evidence in this literature that the rates
quoted are necessarily being achieved.  We show later that it is quite easy, using a seemingly reasonable significance test, to end up with rates that are completely unrepresentative.

\section{Implementation of the general paradigm
\label{sec:ex}}

We described in the previous section three general steps to create a functional topology from multivariate time series data.  In this section, we 1) define our coupling measure, the \cc, 2) develop appropriate significance tests, and 3) integrate these with a technique to account for multiple significance tests.  In Section \ref{sec:res} we apply these specific protocols to simulated and observed data, and show that the choice of the significance test is critical.

\subsection{Step 1: choice of coupling measure}

The choice of coupling measure between pairs of time series 
permits many alternatives \cite{Pereda:2005lr}.  We may select a simple measure of linear coupling (e.g., the cross correlation \cite{Braizer:1973yq, Menon:1996p10970, Gozolchiani:2008p10406, Yamasaki:2008p10391, Kramer:2008p10158} or the coherence \cite{Tharp:1975p10167, Nunez:1997p10972, Towle:1999p10853})
or more sophisticated coupling measures (e.g., \syn\ likelihood \cite{Ponten:2007p4558}, wavelet coherence \cite{Lachaux:2002p10161}, or Granger causality and the related directed transfer function \cite{Kaminski:2001p9200}).  In this manuscript, we choose to focus on a simple measure of coupling based on the \cc.  Although the general statistical hypothesis testing paradigm we adopt here can in principle be applied to any choice of coupling measure, more sophisticated coupling measures may not easily allow for the derivation of computationally tractable significance testing procedures.

Specifically, for a pair of time series $x_i[t]$ and $x_j[t]$, the 
\cc\ at lag $\tau$ is defined as
\be
CC_{ij}[\tau] = \frac{1}{\sigma_i \sigma_j (n-2\tau)} \sum_{t=1}^{n-\tau} (x_{i}[t] - \bar{x}_i)
					(x_j[t+\tau] - \bar{x}_j) \enskip ,
\ee
where $\bar{x}_i$ and $\bar{x}_j$ are the averages, and $\sigma_i$ and $\sigma_j$ are the standard deviations of $x_i[t]$ and $x_j[t]$, respectively, and $n$ is the time series length.  
This quantity can be computed efficiently over a range of  
lags using Fourier transform methods for convolutions.  In our applications,
we first transform the time series at each network node to have
zero mean and unit variance, after which we compute the Fourier transforms
$\tilde{x}_i[\omega]$ and $\tilde{x}_j[\omega]$, multiply the first by the 
complex conjugate of the second, and take the inverse Fourier transform of
the resulting product.  For all of the data considered below, we compute the 
\cc\ for $\tau$ ranging over indices between  $-100$ and $100$ 
(the range of $\tau$ in milliseconds depends on the sampling rate of the data, 
as we describe below).

Our formal measure of coupling will be the {\it maximal} cross correlation i.e.,
$s_{ij}=\max_{\tau} |CC_{ij}[\tau]|$, the maximum of the absolute value of 
$CC_{ij}[\tau]$ over $\tau$.  This measure will serve as our statistic
for testing whether or not to assign an edge between nodes $i$ and $j$,
for each such pair of nodes.

\subsection{Step 2:  Significance test
\label{sec:sig_test}}

Having chosen the test statistic $s_{ij}$, the maximal \cc\ between $\xo$ and $\xt$, do we include a network edge between nodes $i$ and $j$?  To answer this, we will use $s_{ij}$ to test the null hypothesis that $\xo$ and $\xt$ are uncorrelated (i.e., no coupling) against the alternative that they are correlated (i.e., coupling).  Rather
than focusing on testing at a  pre-assigned significance level, we will
instead concentrate first on computing an appropriate \pv\ for
each edge.  Accurate evaluation of the \pv s is critical to successful use of the
false discovery rate principles we employ for the network inference
problem here, as we discuss in Section \ref{sec:fdr}.  We compute the \pv\ in two different ways that make different assumptions about the coupling results.  The first method is an analytic measure and specifically designed for our choice of coupling measure.  The second is more general but computationally expensive.  We define the measures below and, in the next section, apply each to simulated and observed time series data.

\subsubsection{Analytic method
\label{sec:analytic}}

In this section we propose an analytic method.  Frequently such methods involve comparison of a test statistic to a normal distribution.  Following this approach, we would scale $s_{ij}$ by an estimated variance, and then compare this scaled quantity to the standard normal distribution (i.e., with mean 0 and variance 1) to calculate a \pv.
Here, however, this would be naive.

More specifically, under the stated null hypothesis of no coupling the statistic $s_{ij}$ should have mean zero.  A reasonable estimate of the variance of $CC_{ij}[l]$ under the null, motivated by a result of Bartlett~\cite{Bartlett:1946uq, Box:1970kx}, is given by
\be
\label{eq:var}
\widehat{\mathrm{var}}(l) 
  = \frac{1}{n-l} \sum^n_{\tau=-n} CC_{ii}[\tau] CC_{jj}[\tau],
\ee
where the $CC_{kk}[\tau]$ are the autocorrelations of time series $k$ 
at lag $\tau$. This estimate takes a non-trivial form because the \cc\ will depend on the statistical properties of the underlying time series, and in particular on the autocovariance structure.  Spurious \cc s between the two times series are expected even if they are uncoupled~\cite{Netoff:2002fk}, and this variance formula accounts for 
that.

Intuitively we might think to use (\ref{eq:var}) to define $z_{ij} = s_{ij} / \sqrt{\widehat{\mathrm{var}}(\hat{l}_{ij})}$, where $\hat{l}_{ij}$ is the lag corresponding to $s_{ij}$ (i.e., the lag at which the maximum of the absolute value of the \cc\ occurs) and test the significance of the value $z_{ij}$ by comparing it to the standard normal distribution.  Unfortunately, although standardizing $CC_{ij}[l]$ by the estimated variance in (\ref{eq:var}) is sensible for any fixed $l$, use of the standard normal distribution with $z_{ij}$ is not appropriate here, as we explain and illustrate below.

Two potential problems exist in using this naive method to determine 
the significance of $s_{ij}$.  First, the distribution of the cross correlation $C_{ij}[\tau]$
-- strictly speaking -- is normal only in the asymptotic case of large 
sample size $n$.  In finite samples this approximation can be
poor, particularly since the cross correlations are bounded between -1 and 1
while the normal distribution varies over an unbounded range.
Second, we choose $s_{ij}$ as an extremum of the \cc; therefore, we must 
account for this choice when testing the significance of this statistic.  
That is, even in cases where the distributions of the cross correlations
$C_{ij}[\tau]$ are well-approximated by the normal distribution, 
their extrema will not be normally distributed, 
and so \pv s calculated using this distribution will be inaccurate.

To address both of these issues, we propose a more appropriate analytic method: 
the extremum method.  We start by applying the well-known Fisher 
transformation~\cite{fisher1915} to each $CC_{ij}[\tau]$, yielding
\be
\label{eq:extremum_z}
FCC_{ij}[\tau] = \frac{1}{2} \log \frac{1+CC_{ij}[\tau]}{1-CC_{ij}[\tau]}
\enskip ,
\ee
which should more closely follow a normal distribution than the original
$CC_{ij}[\tau]$.  Since this transformation is monotone and symmetric about 
zero, the lag $\hat{l}_{ij}$ maximizing $|CC_{ij}[\tau]|$ will also maximize
$|FCC_{ij}[\tau]|$.  Let $s^F_{ij}$ be the Fisher transformation of 
$s_{ij}$, which we propose to use instead of $s_{ij}$. 

Next, we use results from extreme value theory to approximate the 
distribution of our new test statistic.  We scale the values $FCC_{ij}[\tau]$ 
over $\tau$ by their empirical standard deviations $(\widehat{var}(FCC_{ij}))^{1/2}$ so that the resulting scaled values should approximately follow a standard normal distribution.  If there were no dependency within the time series $x_i[t]$, and instead we observed i.i.d. sequences at each node $i$, then the appropriate standard deviation is known to be $(n-3)^{-1/2}$ \cite{fisher1915}.  But given the dependency in our time series data, we expect that the true standard deviations may differ from this value, and so we choose to estimate them empirically.

The scaled value $z^F_{ij} = s^F_{ij}/(\widehat{var}(FCC_{ij}))^{1/2}$ can be expected 
to behave like the maximum of the absolute values of a sequence of
standard normal random variables.  Using established results for statistics
of this form, we obtain therefore that 
\be
\label{eq:EKTS}
\mathrm{P}[z] \approx \exp(-2\exp(-a_n (z - b_n))) \enskip ,
\ee
where $\mathrm{P}[z] = \Pr\{z^F_{ij}\le z\}$, $a_n = \sqrt{2 \log n}$ and
$b_n = a_n - (2 a_n)^{-1} (\log \log n + \log 4 \pi)$.  A derivation
of (\ref{eq:EKTS}), which holds in the asymptotic sense of large $n$, is provided in Appendix~1.  For the case of $n=201$, as in all of our
numerical results below, $a_n = 3.2568$ and $b_n = 2.6121$.  Using
the approximation above, it is straightforward to calculate
\pv s for the rescaled test statistics $z^F_{ij}$.

Intuitively, the extremum method accounts for our choice of a maximum \cc.
By virtue of the Fisher transformation, the values $FCC_{ij}[\tau]$ will be
approximately normally distributed.  But because we have chosen 
$s^F_{ij}$ as the {\em maximum} of the absolute value of the $FCC$s, 
we expect its value to be skewed towards the tail of the normal 
distribution.  If we had chosen instead any other lag than that 
maximizing the \cc, then the corresponding value $CC_{ij}$ 
(and hence $FCC_{ij}$) would be smaller.  Therefore, our definition of
$s^F_{ij}$ produces \pv s that, if computed from the normal 
distribution, are biased in the sense of being inappropriately small.
From the perspective of our network inference task, this means that --- for 
any given choice of threshold --- we will be more liberal in our assignment of 
edges than we should be.  The distribution in (\ref{eq:EKTS})
essentially corrects for this bias, by explicitly accounting for
our use of the maximum.

\subsubsection{Frequency domain bootstrap method}

The previous method provides an analytic formula for testing the 
significance of $s_{ij}$.  In utilizing 
this formula, we make specific asymptotic distributional assumptions about the test statistic --- the maximal \cc.  These assumptions are likely to be only approximate idealizations of the correlation results emerging from, for example, a complicated physical system like the human brain.  A method to test the significance of $s_{ij}$ that requires fewer assumptions is desirable.  The final method we introduce --- 
the frequency domain bootstrap --- satisfies this desire, but is computationally expensive.

As the name indicates, the method consists of applying the bootstrap 
principle (e.g., \cite{efron}), but in the spectral domain;  methods
of this sort were first proposed in~\cite{franke.hardle}.
We calculate our frequency domain bootstrap through the following steps.
First, we compute the power spectrum (Hanning tapered) of each time series in the network.  We then average these power spectra from all time series, and smooth the resulting average spectrum (moving average of $11$ points).  We use this spectrum estimate ($P[\omega]$) to compute the standardized and whitened residuals for each time series $x_i[t]$:
\be
e_i[t] = \mathrm{iFFT}\big( \tilde{x}_i[\omega] / \sqrt{P[\omega]} \big) 
\enskip .
\ee
Here $\tilde{x}_i[\omega] = \mathrm{FFT}\big(x_i[t]\big)$ is the Fourier
transform of the original time series $x_i[t]$ and
iFFT$( \star )$ is the inverse Fourier transform of $\star$.  
Finally, for each bootstrap replicate, we resample the values
$e_i[t]$ with replacement and compute the surrogate data
\be
\hat{x}_i[t] = \mathrm{iFFT}\big( \tilde{e}_i[\omega] \cdot \sqrt{P[\omega]} \big) \enskip ,
\ee
where $\tilde{e}_i[\omega]$ is the Fourier transform of 
the residuals $e_i[t]$ resampled with replacement.  This last step
ensures that the spectral characteristics (e.g., $1/f^{\alpha}$ behavior)
of the original data are preserved in the surrogate data.

We compute $N_s$ instances of these surrogate data, and for each instance we 
calculate the test statistic $\hat s_{ij}$ for each pair of nodes $i$ and $j$,
(i.e., we calculate the maximum of the absolute value of the \cc\ between the surrogate data $\hat{x}_i[t]$ and $\hat{x}_j[t]$).  The $N_s$ instances of $\hat{s}_{ij}$ 
form a bootstrap distribution of maximum \cc\ values to which we compare $s_{ij}$ observed in the original data and assign a \pv.

Constructing the bootstrap distribution of $\hat{s}_{ij}$ values for all 
node pairs is computationally expensive.  If our network contains $100$ nodes, 
then we would like to compute a bootstrap distribution (and test the 
significance) of each of the $100 \times 99/2 = 4950$ values $s_{ij}$.  
If each bootstrap distribution requires $N_s = 10000$ surrogates, 
a standard choice in the literature, then we construct surrogate data and 
compute the \cc\ over $10^7$ times.  We reduce this expensive operation in 
the following way:  instead of computing bootstrap distributions for all 
electrode pairs, we compute the bootstrap distribution (with $N_s$ surrogates) 
for only a subset of node pairs.  We then define the {\it merged
distribution} as the combined distribution for the entire subset of node pairs. 
We use the merged distribution to test the significance of $s_{ij}$ for all 
node pairs (even pairs not used in calculating the merged distribution).  
Note, however, that in doing so we assume that the null distribution of 
$s_{ij}$ is the same for all node pairs.

\subsection{Step 3:  Control of the false detection rate
\label{sec:fdr}}

To test the statistical significance of the values $s_{ij}$, we may apply either of the two methods described above (or even, practically speaking, the naive method as well).
Networks of, say, $100$ nodes will
consist of $100 \times 99 / 2 = 4950$ $s_{ij}$ values, each with an associated 
\pv.  Clearly, multiple testing is an important concern.  If we simply choose 
a standard \pv\ cutoff for assessing significance (such as $p < 0.05$), 
then we expect the number of network edges incorrectly declared present
to scale proportionally (i.e., roughly $250$ such edges, for an $0.05$ cutoff).
To control for this abundance of false positives, we could define a stricter 
cutoff;  for example, we could use the Bonferroni correction and divide the 
\pv\ threshold by the number of node pairs (i.e., $p < 0.05 / 4950 =  10^{-5}$).
This conservative control of the familywise error rate (the probability of 
making one or more false discoveries) is likely too strict for data in which 
we expect relatively few significant edges {\it a priori,} i.e., for 
sparse networks.

Instead, we employ the less conservative false discovery rate (FDR) to control for multiple testing.  The FDR is defined as the expected proportion of 
erroneously rejected null hypotheses among the rejected 
ones \cite{Benjamini:1995p10138, Reiner:2003p9203}, and various procedures exist for controlling the FDR in practice~\cite{Dudoit:2008uq}.
Generally speaking, the notion of FDR control guarantees that the expected proportion of falsely declared edges in our inferred networks is no more than a pre-specified fraction $q\in (0,1)$.  However, in order for this guarantee to hold, two assumptions 
must be true, namely that (i) statistical $p$-values associated with each test are computed accurately, and (ii) tests are independent.  Of these assumptions, 
the first is critical, while the second is less so.  That the second is
less critical is important in the context of network inference, since the
various tests for declaring presence or absence of edges are clearly
correlated, as they reuse the same time series.  Additionally, if one wishes
to address this dependency, there are extensions of the basic FDR procedure (e.g., see~\cite{Reiner:2003p9203} for useful discussion),
although we do not pursue this here.
On the other hand, inaccurate calculation of $p$-values is known
to be disastrous to FDR principles.  Our analyses presented below
confirm this in the context of our network inference problem, and
the majority of our efforts focus around this point, as we described in Section \ref{sec:sig_test}.

Here we implement the linear step-up FDR controlling procedure of Benjamini and 
Hochberg~\cite{Benjamini:1995p10138}, which is computed 
as follows.  First, order the $m=N(N-1)/2$ \pv s $p_1 \le p_2 \le ... \le p_m$. 
Then, choose a desired FDR level $q$.  Finally, compare each $p_i$ to the 
critical value $q \cdot i / m$ and find the maximum $i$ (call it $k$) 
such that $p_k \le q \cdot k / m$ (and therefore $p_{k+1} > q \cdot (k+1) / m$).
We reject the null hypothesis that time series $\xo$ and $\xt$ are uncoupled 
for $p_1 \le ... \le p_{k}$.

The choice of $q$ determines a threshold \pv, for which we 
declare all features significant up through that threshold~\cite{Storey:2003gf}.
The value $q$ represents an upper bound on the expected proportion of false 
positives among all declared edges in our inferred network (i.e., among all
node pairs for which $s_{ij}$ was declared to be significant.)
For example, if we fix $q=0.05$ and find $100$ significant values of $s_{ij}$, 
then we expect $0.05 \cdot 100 = 5$ false positives (i.e., five false edges 
in the $100$ edge network).

\section{Results
\label{sec:res}}
We analyze three data sets using the procedure defined in Section \ref{sec:ex}. Two data sets we create with specific (known) structural topologies, to which we compare the functional topologies extracted through analysis of the dynamic data. The third we observe from a human epileptic subject undergoing invasive electrical monitoring of the cortex during seizure.

\subsection{Pink noise data}
\begin{figure}[tbh]
\centering
\includegraphics[angle=0, height=5.0cm, width=8.6cm]{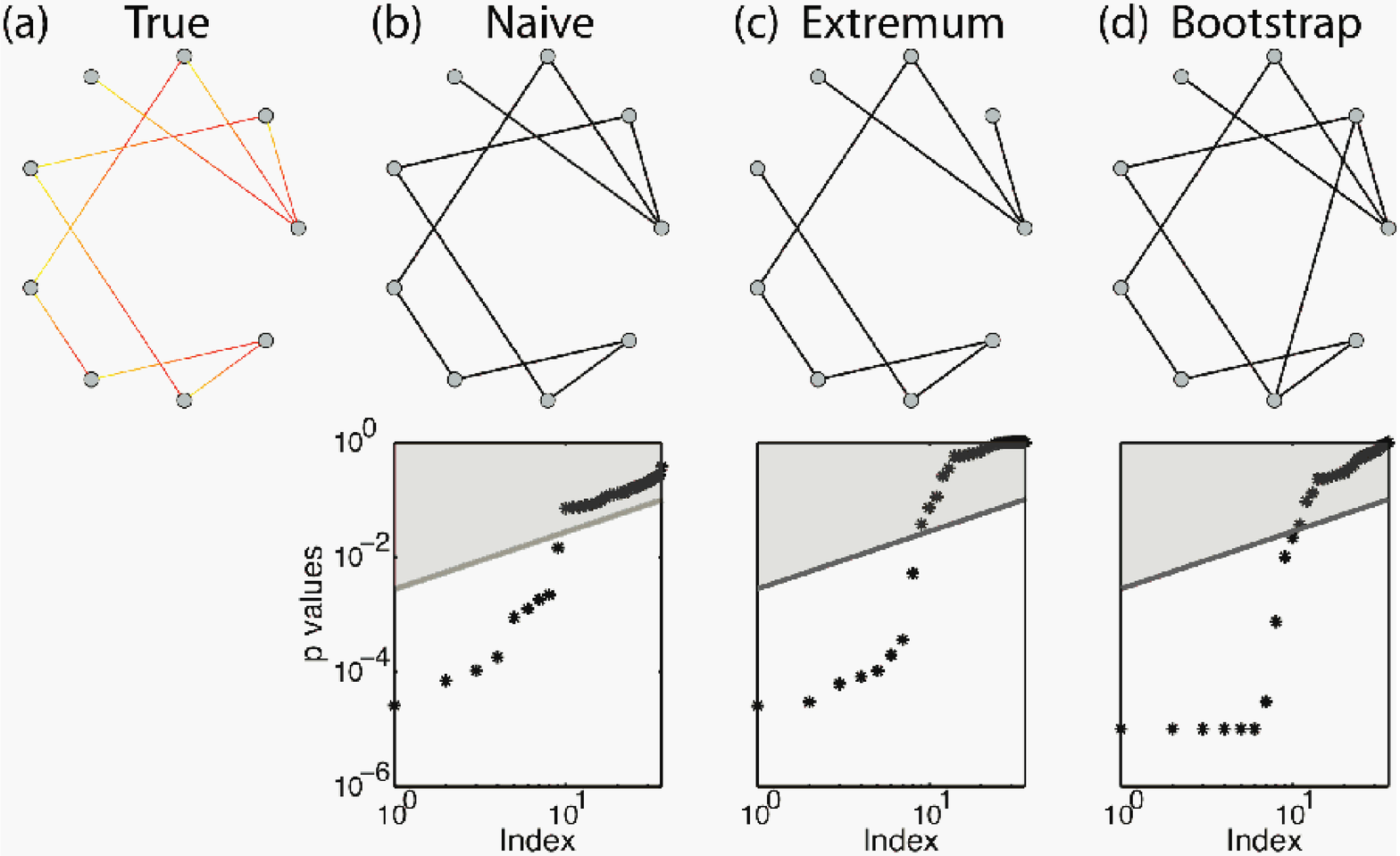}
  \caption{\label{fig:9node}  (Color online)  For the colored noise network all three measures perform equally well and detect the underlying structural topology.  In the upper row, each circle indicates a node.  (a)  The true connectivity of the network.  With colored (or shaded) lines we indicate directed connections between nodes;  connections initiate from the red (dark) line end and terminate at the yellow (light) line end.  (b-d)  The \pv s (lower) and corresponding functional network topologies (upper) derived from the naive method (b), extremum method (c), and bootstrap method (d).  The dark gray line in the lower figures indicates the threshold for the linear step-up FDR procedure;  we consider \pv s below this line --- in the unshaded region --- significant.  All three significance tests capture the functional topology equally well.}
\end{figure}

Many time series data produced by natural systems possesses a $1/f^{\alpha}$ power spectrum \cite{Dutta:1981p10173, Gisiger:2001p10174, LinkenkaerHansen:2001p10169}.  To mimic this behavior, we create a nine-node network, first by generating $500$ ms (sampling interval $1$ ms) of independent colored noise ($\alpha = 0.33$) data $w_i[t]$ at each node $i$.  We then connect node $i$ to $j$ by adding pointwise to $w_j[t]$ the data $w_i[t]$ scaled by a factor of $0.4$.  For example, we connect node \#1 to \#2 by adding to $w_2[t]$ the time series $0.4 \cdot w_1[t]$ for each time point $t$ to create $x_2[t] = w_2[t] + 0.4 \cdot w_1[t]$, the time series associated with node \#2.  In Fig.\ \ref{fig:9node}(a) we illustrate the topology of the constructed network;  a total of nine directed connections exist.

Having established the network topology, we now attempt to recover it directly from the time series data.  To do so, we apply our coupling measure ($s_{ij}$, the maximum of the absolute value of the \cc) pairwise to all $m = 9 \cdot 8 / 2 = 36$ electrode pairs in the network.  We then test the significance of each $s_{ij}$ and compute a \pv\ using the analytic and computational procedures defined above.  We begin with the naive method, whose \pv s we plot as asterisks in the lower portion of Fig.\ \ref{fig:9node}(b).  Plotted in increasing order, these \pv s range from $\sim 10^{-5}$ to $0.3$.  We fix $q=0.10$ and also plot in the lower portion of Fig.\ \ref{fig:9node}(b) the line of slope $q/m = 0.10/36 = 0.0028$ and zero intercept.  Following the linear step-up FDR procedure, we reject the null hypothesis of no coupling for those (nine) electrode pairs with \pv s below this line.  We plot in the upper portion of Fig.\ \ref{fig:9node}(b) the (nine) ``significant edges" corresponding to the significant \pv s.  Our confidence in this nine node network --- derived from the time series data --- is high;  from the FDR procedure we expect $0.10 \cdot 9 \sim 1$ false positive edge (i.e., one spurious edge between uncoupled nodes).  In this case, we find exact agreement between the known network topology (Fig.\ \ref{fig:9node}(a)) and the derived topology.  We note that, for sake of clarity, we chose a simple coupling measure that does not determine edge direction.  More sophisticated coupling measures that indicate edge direction may be employed following the general paradigm outlined above, as we discuss in Section \ref{sec:dis}. 

In Figs.\ \ref{fig:9node}(c) and \ref{fig:9node}(d), we show the topology derived using the extremum and bootstrap methods, respectively.  In both cases, we follow the linear step-up FDR procedure with $q=0.10$ to identify significant edges.  For the extremum method (Fig.\ \ref{fig:9node}(c)) we identify eight significant edges, one less than expected.  We compute our confidence in the network using the FDR procedure and anticipate $0.10 \cdot 8 \sim 1$ false positive edge.

To compute the frequency domain bootstrap, we first calculate the average power spectrum of all (nine) nodes.  We then create a merged distribution using a subset of ten electrode pairs (of the possible $36$) and $N_s = 10000$ for each surrogate distribution.  The resulting merged distribution contains $10 \cdot N_s = 10^5$ points;  therefore, the smallest \pv\ we can compute through this method is $10^{-5}$.  We find, in this case, six \pv s at this detection limit.  Using the bootstrap method (Fig.\ \ref{fig:9node}(d)), we identify ten network edges, one more than expected.  We do expect $0.10 \cdot 10 = 1$ false positive edge in the network, although given only the time series data, we could not identify which of the ten edges is spurious.

These simulation results suggest that all three measures of edge significance perform equally well.  This is surprising, especially for the naive method in which we neither Fisher transform the maximal correlation values (to induce normality), nor account for our choice of an extremum (the maximum of the absolute value of the \cc).   The naive method succeeds, in this case, because the two omissions appear to balance.
Omitting the Fisher transformation increases the \pv s we observe, while utilizing the normal distribution with zero mean --- not the extremum distribution --- decreases the \pv s.  One omission compensates the other so that, in this case, the resulting \pv s are approximately correct.  Unfortunately, we cannot rely on this delicate balance to always succeed as we illustrate in the next example.

\subsection{Simulated neural data}
\begin{figure}[tbh]
\centering
\includegraphics[angle=0, height=5.0cm, width=8.6cm]{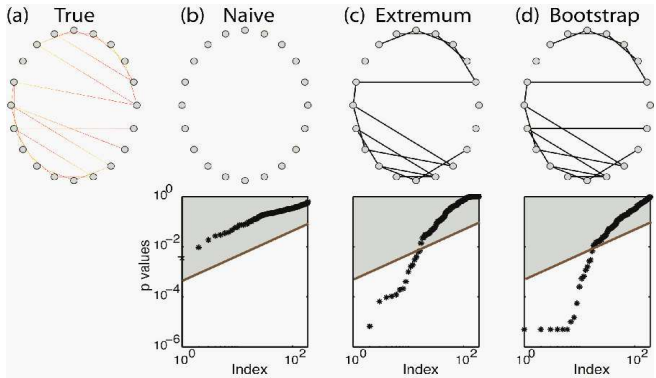}
  \caption{\label{fig:spike_net}  (Color online)  For the simulated neural data the choice of statistical test is vital to construct an appropriate network.  (a) The data consist of twenty cell groups (gray circles) and $22$ connections between cell groups.  Directed connections proceed from the red (dark) to yellow (light) end of each line.  (b-d) The functional networks deduced.
The naive method (b) identifies no significant \pv s;  with $q=0.10$ in the linear step-up FDR procedure, none of the \pv s lie below the (dark gray) line $(q/m) \cdot i$.
The extremum method (c) identifies $17$ significant edges (of which we expect two are false positives);  $14$ match the structural network in (a).
The bootstrap method (d) detects $18$ edges, of which we expect $2$ false positives. This procedure detects $15$ (of the $22$) true edges.}   
\end{figure}

In the previous model, we simulated colored noise activity possessing a $1/f^{\alpha}$ falloff of the power spectrum.  We now consider a more realistic model of interacting neural populations.  We provide a brief description of the model here;  more details may be found in Appendix 2.  The model consists of $1000$ neurons divided into twenty groups of $50$ cells.  Within each group we include strong connections (\ex\ synapses) between randomly chosen neurons;  activity initiated by a few neurons in a group quickly spreads to the other neurons of the same group.  Between cell groups, we establish only weak (\ex\ synaptic) connections joining individual neurons of specific cell groups.  We illustrate the topology of these weak connections between cell groups in Fig.\ \ref{fig:spike_net}(a).  In this figure, each gray circle represents a cell group (of $50$ strongly connected neurons) and lines represent connections between cell groups.  With this connectivity in place, we simulate the neural dynamics and compute the average population activity of each group.  We then employ the general paradigm described above to the resulting neural activities and compare the measured functional connectivity (i.e., the pattern of connectivity inferred from the simulated neural dynamics) to the known structural connectivity between nodes shown in Fig.\ \ref{fig:spike_net}(a).  The results, as we show below, depend upon the significance test we employ.

We apply the coupling measure pairwise to all $m = 20 \cdot 19 / 2 = 190$ possible group pairs in the network and test the significance of each result by computing a \pv\ using one of the three procedures defined above.  We begin with the naive method, whose \pv s we plot as asterisks in the lower portion of Fig.\ \ref{fig:spike_net}(b).  With $q=0.10$ in the linear step-up FDR procedure, we find no significant values of maximal \cc;  none of the \pv s lie below the line $(q/m) \cdot i$.  The resulting (trivial) network --- shown in the upper portion of Fig.\ \ref{fig:spike_net}(b) --- contains no edges.  The other two significance tests produce nontrivial networks.  Using the extremum method and linear step-up FDR procedure (with $q=0.10$) we identify $17$ significant edges.  The resulting network, shown in Fig.\ \ref{fig:spike_net}(c), correctly identifies $14$ edges and possesses three erroneous edges (i.e., edges we identify in the functional network that do not exist in the structural network).  We expect from the FDR procedure $q \cdot 17 \sim 2$ false positives, in approximate agreement with the three erroneous edges observed.  Finally, we show in Fig.\ \ref{fig:spike_net}(d) the \pv s and network determined using the bootstrap method.  In this case, we detect $18$ edges (and expect $2$ false positives).  This procedure detects $15$ (of the $22$) true structural edges and produces three erroneous edges, again in approximate agreement with the number of false positives expected.

In all three cases, the functional topology derived from the mean dynamics fails to capture exactly the true structural topology of the network.  The naive method detects no significant edges and performs most poorly.  This is not surprising;  we expect that the un-normalized \pv s and incorrect distribution of maximal correlation values will compromise the naive method.  The extremum and bootstrap methods produce similar functional networks that approximate the true structural network.  Although both measures make mistakes, the FDR procedure provides an estimate for the number of erroneous edges to expect.  We conclude that, for these simulated data, the extremum and bootstrap methods outperform the naive method and qualitatively reproduce many (but not all) of the network edges.

\subsection{Human ECoG data
\label{sec:human}}
In the previous two examples, we applied the coupling analysis to networks with known structural topology.  This allowed us to compare the derived functional topology with the true structural topology and determine each method's performance.  As a last illustration of the methods, we consider voltage activity recorded directly from the cortical surface (electrocorticogram or ECoG data) of an epileptic human subject for clinical purposes (Appendix 3).  We focus on a short interval (1 s) of data recorded from $97$ electrodes while the subject experienced a seizure.  We apply all three methods to the data and compare the resulting (functional) networks.  In this case, the structural connectivity is unknown.  We find that, as before, the extremum and bootstrap methods produce consistent results.

\begin{figure}[tbh]
\centering
\includegraphics[angle=0, height=5.0cm, width=8.6cm]{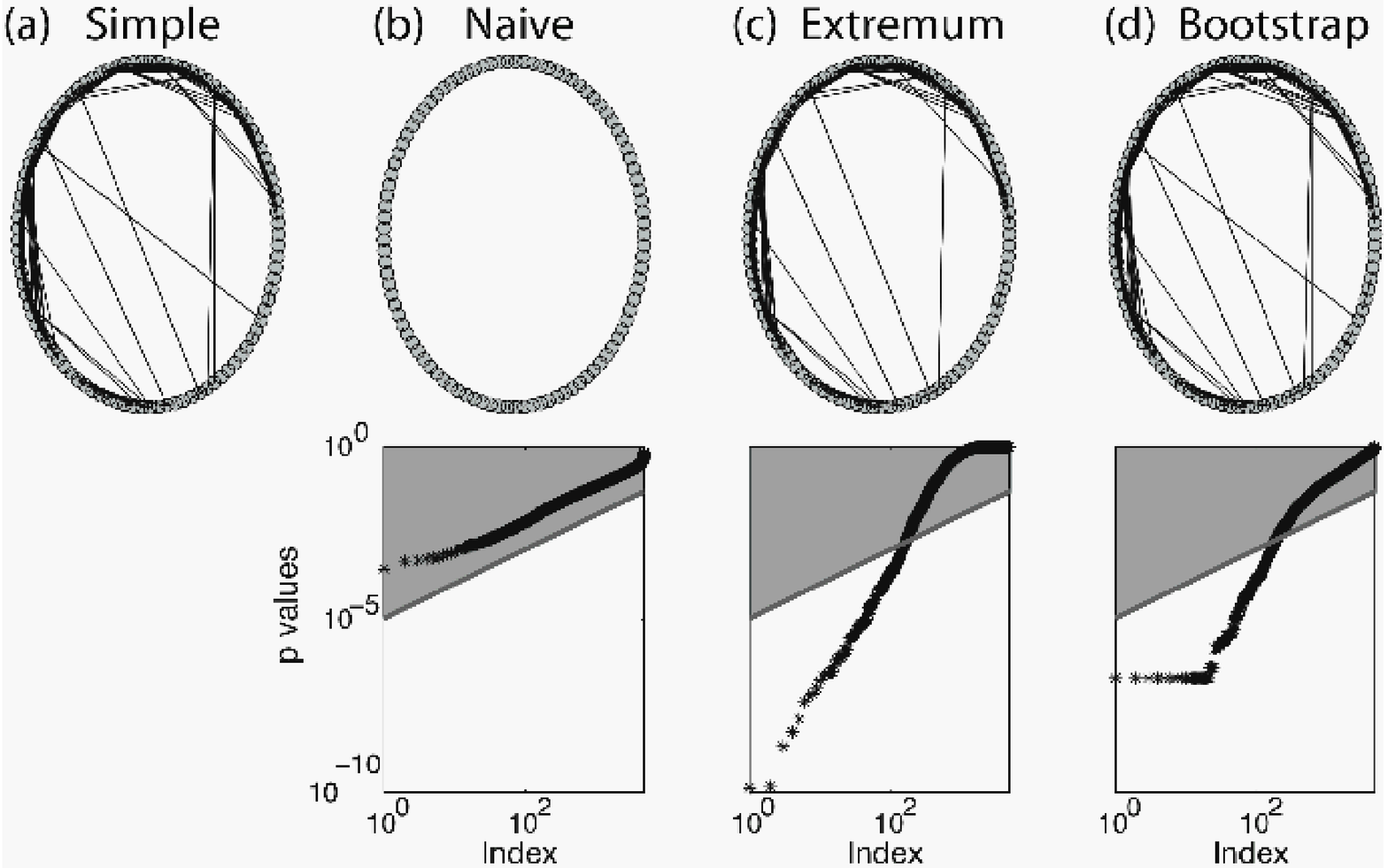}
\caption{\label{fig:human_sz_net}  Functional networks constructed from $1$ s of ECoG data recorded at $97$ electrodes during a seizure depend upon the statistical test we perform.  (a) A simple threshold network with edges (black lines) drawn between nodes pairs exhibiting sufficient functional coupling, $s_{ij} > 0.75$.  (b-d, lower)  The $4656$ \pv s calculated from the naive method (b), extremum method (c), and bootstrap method (d).  For each method, we fix $q=0.05$ in the FDR procedure. (b-d, upper)  The corresponding functional networks.  The naive method (b) detects no significant edges and the corresponding network is trivial.  The network created from the extremum method (c) contains $162$ edges, and from the bootstrap method (d) $187$ edges.}
\end{figure}

We show the deduced functional networks in Figs.\ \ref{fig:human_sz_net}(b-d).  In each case, we  test the significance of $m=97 \cdot 96 / 2 = 4656$ maximal \cc\ values, and use a linear step-up FDR procedure with $q=0.05$ to define significant \pv s.  For the naive method (Fig.\ \ref{fig:human_sz_net}(b)) we find no significant \pv s and the corresponding trivial network contains no edges.  We note that the node locations in Fig.\ \ref{fig:human_sz_net} do not correspond to their physical locations on the human cortex.  Instead, we simply arrange the nodes in a circle. 

From the extremum and bootstrap methods we create similar networks.  For the former, we identify $162$ significant edges (of which we expect $9$ false positives) as drawn in Fig.\ \ref{fig:human_sz_net}(c).  For the latter, we select $500$ electrode pairs (of the possible $4656$ pairs) to compute surrogate distributions, each distribution containing $N_s = 10000$ realizations.  The smallest \pv\ detectable in the resulting merged distribution is $2 \times 10^{-7}$.  Using this method we find the $187$ significant edges drawn in Fig.\ \ref{fig:human_sz_net}(d), of which we expect $10$ false positives.

Comparing the functional networks deduced from the extremum and bootstrap methods, we find that the two are similar.  Moreover, we show in Fig.\ \ref{fig:human_sz_net}(a) a fourth functional network constructed using a simple threshold procedure;  we include edges only between those node pairs with $s_{ij} > 0.75$.  Remarkably, all three networks are qualitatively similar although we use different techniques to construct each network.  Of course the simple threshold network does not indicate our confidence in the network:  how many edges in Fig.\ \ref{fig:human_sz_net}(a) are false positives?  In addition, we note that the bootstrap method is computationally expensive;  constructing the surrogate distribution requires approximately $90$ minutes on a $2$ GHz Core Duo processor and therefore at least $45$ hours to construct the networks for $30$ s of seizing activity.  The extremum method, designed for our particular choice of coupling measure, identifies a network similar to the bootstrap method in a computationally efficient way.

\subsection{Human ECoG data:  shuffled}

For the human ECoG data, we do not know the structural network (i.e., we do not know the topology of chemical and electrical connections between neurons in these cortical regions).  Therefore, we cannot validate the functional networks shown in Fig.\ \ref{fig:human_sz_net} by comparison with anatomical connections.  However, we can manipulate the ECoG data to disrupt functional connections and verify that our significance tests detect no coupling.  To do so, we create a new data set:  we assign to each electrode $1$ s of data chosen at random from a $120$ s interval that includes $60$ s of pre-seizure and $60$ s of seizure activity.  For example, electrode \#1 may contain ECoG data from $t=[8.2,9.2]$, electrode \#2 data from $t=[97.0, 98.0]$, electrode \#3 from $t=[110.4, 111.4]$, and so on.  With the data chosen in this way, we expect only weak associations between electrode pairs.

\begin{figure}[tbh]
\centering
\includegraphics[angle=0, height=5.0cm, width=8.6cm]{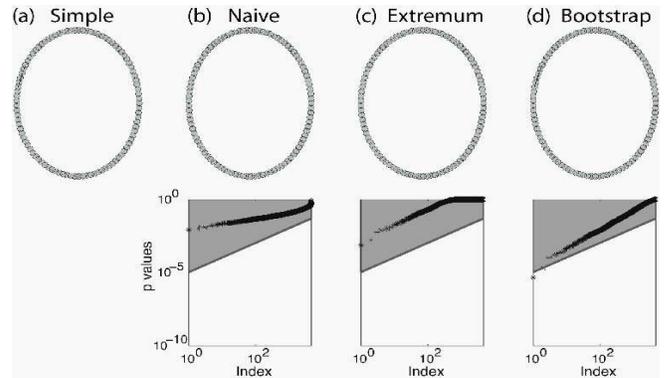}
\caption{\label{fig:human_scram_net}  By shuffling the ECoG data, we eliminate coupling between the time series and detect no (or few) edges.  (a) A simple threshold network with edges drawn between node pairs with sufficient functional coupling ($s_{ij} > 0.75$) detects two edges located at the left of the network.  (b-c)  The \pv s\ (lower) and corresponding networks (upper) derived from the (b) naive, (c) extremum, and (d) bootstrap methods.  Only the latter detects two edges (of which we expect $1$ false positive).}
\end{figure}

We follow the procedure described above to analyze these ``shuffled" data.  We compute the maximal \cc\ for each electrode pair, and show the corresponding \pv s and functional networks in Fig.\ \ref{fig:human_scram_net}(b-c).  With $q=0.05$, we find no significant coupling using the naive or extremum methods.  We do detect $2$ significant edges with the bootstrap method (of which we expect $1$ false positive).  These significant edges match those determined using a simple threshold procedure ($s_{ij} > 0.75$) whose network we show in Fig.\ \ref{fig:human_scram_net}(a).  We conclude that the three significance tests behave as expected for the shuffled data;  if we disrupt the coupling in the data, we expect trivial functional networks.

\section{Discussion
\label{sec:dis}}

Our increased ability to collect multivariate spatiotemporal data (e.g., from high density electrode arrays) necessitates the construction and analysis of complex functional networks.  In this manuscript, we adopted a statistical hypothesis testing paradigm for constructing such functional networks.  This paradigm involved three steps:  1) choice of an association measure, 2) definition of a significance test, and 3) accounting for multiple significance tests.  Although the paradigm itself is quite general, the details accompanying each step are problem specific.

Here we developed this general paradigm for multivariate time series data.  For the association measure we chose the maximum of the absolute value of the \cc.  We defined two approaches to significance testing (one analytic and the other computational), and employed a linear step-up FDR procedure to account for multiple tests.  Applying these techniques to three data sets, we showed that the choice of significance test was critical.  Without accurate \pv s for each network edge, we lack confidence in the resulting network.

The general paradigm outlined in Section \ref{sec:general} applies to any choice of association measure.  In this work we focused on this simple \cc\ measure for two reasons.  First, the measure is computationally efficient.  Second, analytic expressions exist (or can be derived) to test the significance of each \cc\ result.  More appropriate coupling measures exist \cite{Pereda:2005lr} that may perhaps improve the network results we present here.  In particular, measures that distinguish direct from indirect interactions and incorporate the flow of information \cite{Seth:2005p10577, Schelter:2006p9191} would be of use.  However, choosing a more sophisticated association measure does not guarantee more accurate functional networks.  The coupling measure must also include an accurate significance test;  without precise \pv s for each network edge, we weaken our measures of network confidence.

Researchers in various other contexts have followed a similar strategy of associating \pv s with each network edge and accounting for multiple significance tests (e.g., \cite{Achard:2006p4675,Valencia:2008p9750}).  Our numerical results illustrate how the choice of an appropriate significance test associated with a specific coupling measure is critical. That a measure possesses a significance test does not guarantee accurate \pv s;  typically significance tests make specific assumptions about the data.  For example, we found that the naive method --- although perhaps intuitively appealing --- was inappropriate because we did not account for taking the {\it maximum} of the absolute value of the \cc, and thus produced inaccurate \pv s and inaccurate networks.  Therefore, we utilized two additional, complimentary measures.   By testing the paradigm on simulated data with known physical connectivity we deduced appropriate significance tests for the association measure implemented here.

We note that trivial networks (e.g., networks without edges as in Fig.\ \ref{fig:human_scram_net}) rarely appear in practice.  Upon finding a trivial network, a common response is to adjust the network threshold to include more edges, perhaps until the network becomes strongly connected.  To follow a similar strategy here we increase the value of $q$ in the linear step-up FDR procedure.  If, for example, we set $q=0.5$ (instead of $q=0.05$) we may detect new significant network edges.  But by increasing $q$ we decrease our confidence in the network;  with $q=0.5$, we expect half of the network edges declared significant are false positives.  Thus, through our choice of $q$, we balance the number of edges detected with our confidence in the network.

The typical approach to construct functional networks from multivariate time series data involves thresholding an association measure.  For example, we may define edges between nodes whose maximal \cc\ exceeds $0.75$, as in \cite{Kramer:2008p10158}.  This procedure for constructing a network suffers from numerous inadequacies. First, we lack a measure of confidence in the resulting network.  With this choice of $0.75$ as threshold how many spurious edges do we expect, and does this number change as we vary the threshold?  Second, we expect the choice of threshold may depend on the particular instance of data observed.  For example, in constructing functional networks of ECoG data recorded during seizure, the threshold may vary from patient to patient, depending on mechanisms intrinsic to each individual.  Finally, a more robust approach to constructing functional networks must propagate error in the association measure to uncertainty in network measures (e.g., to uncertainty in measures of degree or betweenness).

We propose that choosing a threshold value of $q$, rather than a threshold value of an association measure, constitutes a more rigorous procedure for establishing functional networks.  By choosing the threshold through the use of formal statistical hypothesis tests, we create functional networks with specified levels of network uncertainty that may be calibrated across a population of multivariate data.  In the future, we will use this approach to study how uncertainty in the association measure affects uncertainty in network characteristics, and how to adopt these procedures for weighted (rather than binary) networks.
Combined with biophysical models, robust techniques to create functional networks will perhaps illuminate the mechanisms that produce the observed activity and, when necessary, suggest how to alter this activity.

\acknowledgments{MAK holds a Career Award at the Scientific Interface from the Burroughs Wellcome Fund.  UTE was supported by NSF-IIS-0643995.  SSC was supported by the Rappaport Foundation and the Executive Committee of Research of Partners Healthcare.  EDK was supported by ONR award N00014-06-1-0096.}

\section{Appendix 1: Derivation of (\ref{eq:EKTS}).}

Suppose that $Z_1,\ldots,Z_n$ are independent and identically distributed
normal random variables, with mean 0 and variance 1.  Define
$M_n = \max_i(Z_i)$ and $m_n = \min_i(Z_i)$.  Then
\begin{eqnarray*}
  \Pr\left( \max_i |Z_i| \le z\right) & = & 
  \Pr\left( M_n \le z\quad ,\quad
	    m_n \ge -z\right) \\
 & \approx & \Pr\left(M_n \le z\right)\, 
       \Pr\left(m_n \ge -z\right) \\
 & = & \left[ \Pr\left( M_n \le z\right) \right]^2 \enskip .
\end{eqnarray*}
The approximate equality in the second line follows by asymptotic 
independence of the max and min (e.g., Theorem 1.8.2 of \cite{leadbetter.etal}), while the equality in the last line follows by symmetry of the normal
distribution.  Now by, for example, Theorem 1.5.3 of \cite{leadbetter.etal}
we have
\be
\Pr\left(a_n(M_n - b_n)\le z\right) \approx \exp\left(-e^{-z}\right) \enskip ,
\ee
with equality holding asymptotically in $n$.  As a result, we obtain the 
expression in (\ref{eq:EKTS}) in the ideal case that the standardized statistic
$z^F_{ij}$ derives from cross correlations $CC_{ij}[\tau]$ that are
independent.  Although these cross correlations will clearly be dependent,
the approximation (\ref{eq:EKTS}) nevertheless can be expected to hold
fairly generally, as the basic limiting extreme value distribution 
used here is quite robust for sequences of normal random variables
under a range of dependency conditions, for both stationary and
even non-stationary cases.  See, for example, Chapters 4 and 6 of
\cite{leadbetter.etal}.

\section{Appendix 2:  Neural model}
\begin{figure}[tbh]
\centering
\includegraphics[angle=0, height=2.5cm, width=8.6cm]{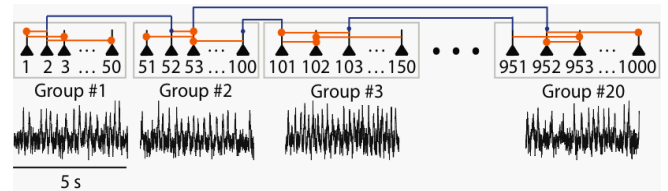}
  \caption{\label{fig:cartoon_net}  (Color online)  A cartoon representation of the network used to simulate the neural data.  Each group of neurons contains $50$ cells, represented as filled triangles.  Within a group, we include many, strong synaptic connections (terminating at large circles, orange).  Between groups, we include few, weak synaptic connections (terminating at small circles, blue).  We show examples of the average activity of each group (from which we construct the functional networks) in the lower portion of the figure.}   
\end{figure}

We model the dynamics of each neuron with two ordinary differential equations, one to represent the membrane voltage, and the other a membrane recovery variable \cite{Izhikevich:2003ve}.  We choose the model parameters so that each neuron generates regular spiking activity (i.e., $a=0.02$, $b=0.2$, $c=-65.0$, $d=8.0$ in \cite{Izhikevich:2003ve}).  We then connect the neurons with \ex\ synaptic connections to establish two connectivity patterns:  strong-local connectivity and weak-distant connectivity.  In both cases, we divide the neurons into groups of $50$ cells numbered sequentially (i.e., group \#1 contains cells $\{1-50\}$, group \#2 cells $\{51-100\}$, group \#3 cells $\{101-150\}$, and so on.)  Within each local group of cells, we create $1200$ directed \ex\ synapses (of the possible $50 \times 49 = 2450$ directed pairs with no self synapses).  Each synapse is assigned a uniform random conduction delay between $0$ and $10$ ms and synaptic strength chosen uniformly between $0$ and $15$.  These synapses establish the strong-local connectivity within a cell group and define the twenty cell groups in the network;  see Fig.\ \ref{fig:cartoon_net}.

We also create weaker synaptic connections between the local cell populations.  To do so, we select two groups (e.g., \#1 and \#8) and create $550$ \ex\ synapses from neurons in one group (e.g., \#1) to neurons in another (e.g., \#8).  These ``distant" synapses are weaker than the local connections;  we assign the synaptic strengths smaller random values (chosen uniformly between $0$ and $5$) and uniform random conduction delay between $0$ and $10$ ms.  We illustrate the $22$ distant connections between the twenty cell groups in Fig.\ \ref{fig:spike_net}(a).  Each gray circle represents a local cell group (i.e., a subset of $50$ neurons).  The colored (shaded) lines represent the distant synaptic connections between groups.  In addition to the local and distant synaptic inputs, we also include strong synaptic input (from the thalamus, say) to one randomly chosen neuron each millisecond, causing this neuron to generate an action potential.  We follow the algorithm in \cite{Izhikevich:2003ve} to simulate the neural population for $5000$ steps (or $5$ s) with a sampling interval of $1$ ms.  The model is similar to recent simulations \cite{Izhikevich:2006p2930,Lubenov:2008p7730}, except that we introduce here connectivity with a particular structural topology.

In the human ECoG recordings described in Section \ref{sec:human}, we observe the dynamics of postsynaptic potentials produced by large neural populations, not the spiking activity of individual neurons \cite{Mitzdorf:1985p2948, Lakatos:2005lr}.  To mimic these population dynamics, we construct the mean activity of the local neural groups in the following way.  First, we define $I[t]$ as the total current input to each neuron at time $t$.  In the model equations we simulate here, these current inputs change instantaneously \cite{Izhikevich:2003ve}.  In reality, current inputs follow the opening and closing of channels and evolve more slowly \cite{Koch:1999fk}.  To approximate these slow dynamics at postsynaptic neuron $j$, we use the following equation:
\be
\dot{s_j} = I_j[t] (1-s_j) - s_j / 20,
\ee
where $s_j$ represents the state of the synapse at neuron $j$, and $I_j[t]$ represents the total \ex\ input current to the neuron $j$ at time $t$.  We note that, for simplicity, we approximate the total synaptic input to neuron $j$ as a single synapse with dynamics driven by $I_j[t]$, the activity of all neurons presynaptic to $j$.  When $I_j[t]$ is large, \ex\ current enters neuron $j$ and $s_j \rightarrow 1$.  When $I_j[t]$ is small, $s_j \rightarrow 0$ with a decay time constant of $20$ ms, and neuron $j$ approaches its resting potential.  We define the mean activity of a neural group as the average of $s_j$ over the local (fifty) cell population.  For example, we compute the mean activity of population \#1 as the average value of $s_j$ for neurons $j=\{1, 2, ..., 50\}$.  We only use $s_j$ to define the mean population activity;  these synaptic dynamics do not impact the voltage dynamics of the model neurons.

With the population activity defined in this way (i.e., as the mean synaptic dynamics within a neural group), we simulate $5$ s of neural dynamics and record the average activity of each group.  We then scale the group activity to have zero mean and unit variance and add Gaussian noise (zero mean and $0.55$ variance) to each sample of each time series.  Finally, we downsample the group traces by a factor of five, reducing the sampling rate (from 1000 Hz to 200 Hz) to decease subsequent computational time;  we show examples of the resulting time series data in Fig.\ \ref{fig:cartoon_net}.

\section{Appendix 3:  Human subject data}
The ECoG data were recorded from a 37 year old male patient with medically refractory epilepsy whose seizures began at age 3.  Following the failure of seven antiseizure medications and a vagal nerve stimulator, and upon recommendation of his clinical team consisting of epileptologists and neurosurgeons, the decision was made to pursue resection of the tissue from which the seizures arose.  To this end, subdural grids of electrodes were implanted.  The goal of this procedure was to identify the epileptogenic zone --- the region of the brain producing recurrent seizures --- and surgically remove it \cite{Rosenow:2001p1837}.

The ECoG recordings consisted of $100$ electrodes placed directly on the cortical surface ($92$ electrodes over the left frontal and temporal lobes) and within deep brain regions ($8$ electrodes within the temporal lobe).  Following electrode implantation, the subject was admitted to a specialized monitoring unit and data recorded continuously at 500 Hz for ten days.  During this time, four seizures were observed;  to illustrate the measures, we analyze only the second seizure here.  Analysis of the collected data was approved through the Partners Health Care Human Research Committee and the Charles River Campus Institutional Review Board.

\begin{figure}[tbh]
\centering
\includegraphics[angle=0, height=5.0cm, width=8.6cm]{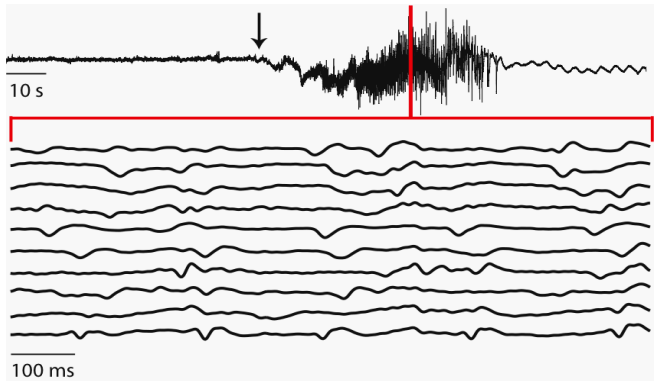}
\caption{\label{fig:human_traces}  (Color online)  Examples of the ECoG data recorded from the human subject during seizure.  In the upper trace we show $160$ s of data recorded from a single electrode;  the seizure begins at the transition from low amplitude to high amplitude fluctuations denoted by the arrow.  We indicate the one second interval analyzed with the vertical red line.  The lower ten traces illustrate the voltage activity recorded from ten (of the 97) electrodes during this one second interval;  we apply the coupling measure to all pairs of these data.}
\end{figure}

We apply our coupling analysis to simultaneous recordings from $97$ electrodes;  three electrodes --- suffering from extreme artifacts --- were discarded.  Before beginning the coupling analysis, we process the ECoG data in the following way.  First, we lowpass filter the data (two-way least-squares FIR filtering) below $55$ Hz to isolate the low frequency components.  We therefore ignore higher frequency activity that may delineate seizure onset \cite{Bragin:1999p503, Traub:2001fk} and instead focus on the high amplitude, low frequency oscillations that characterize unequivocal clinical seizures \cite{Litt:2002fk}.  We then choose a $1$ second interval of the ECoG data during the seizure.  We chose this short interval to balance two competing needs:  stationarity and sufficient data.  For the former, we must choose an interval in which the voltage dynamics at each electrode remain approximately consistent (i.e., exhibit oscillations of the same approximate character).  For the latter, we must choose an interval that contains enough data to calculate the coupling measure (e.g., a $50$ ms interval would fail to capture some slow oscillations characteristic of a seizure).  Finally, we normalize the data from each electrode within the $1$ second interval to have zero mean and unit variance.  We show examples of the ECoG data employed in the analysis in Fig.\ \ref{fig:human_traces}.


\end{document}